\documentclass[3p,preprint,twocolumn]{elsarticle}
\journal{Surface Science}

\usepackage{graphicx}
\usepackage{upgreek,mathtools}
\usepackage{hyperref}
\usepackage[caption=false,labelformat=parens]{subfig}
\newcommand{\GM}{$\overline{\Gamma \text{M}}$}
\newcommand{\GK}{$\overline{\Gamma \text{K}}$}

\begin{document}

\begin{frontmatter}
\title{A Helium-Surface Interaction Potential of Bi$_2$Te$_3$(111)\\ from Ultrahigh-Resolution Spin-Echo Measurements}

\author[CAM,GRZ]{Anton Tamt\"ogl\corref{cor}}\ead{tamtoegl@gmail.com}
\author[GRZ]{Michael Pusterhofer}
\author[ARH1]{Martin Bremholm}
\author[ARH1]{Ellen M. J. Hedegaard}
\author[ARH1]{Bo B. Iversen}
\author[ARH2]{Philip Hofmann}
\author[CAM]{John Ellis}
\author[CAM]{William Allison}
\author[MAD]{S. Miret-Art\'es}
\author[GRZ]{Wolfgang E. Ernst}

\cortext[cor]{Corresponding author. Tel +43 316 873 8143}
\address[GRZ]{Institute of Experimental Physics, Graz University of Technology, Petersgasse 16, 8010 Graz, Austria}
\address[CAM]{Cavendish Laboratory, University of Cambridge, J. J. Thomson Avenue, CB3 0HE, Cambridge, United Kingdom.}
\address[ARH1]{Center for Materials Crystallography, Department of Chemistry and iNANO, Aarhus University, Denmark}
\address[ARH2]{Department of Physics and Astronomy, Interdisciplinary Nanoscience Center (iNANO), Aarhus University, Denmark}
\address[MAD]{Instituto de F\'isica Fundamental (IFF-CSIC), Serrano 123, 28006 Madrid, Spain.}
%%%%%%%%%%%%%%%%%%%%%%%%%%%%%%%%%%%%%%%%%%%%%%%%%%%%%%%%%%%%%%%%%%%%%


\begin{abstract}
We have determined an atom-surface interaction potential for the He--Bi$_2$Te$_3$(111) system by analysing ultrahigh resolution measurements of selective adsorption resonances. The experimental measurements were obtained using $^3$He spin-echo spectrometry. Following an initial free-particle model analysis, we use elastic close-coupling calculations to obtain a three-dimensional potential. The three-dimensional potential is then further refined based on the experimental data set, giving rise to an optimised potential which fully reproduces the experimental data. Based on this analysis, the He--Bi$_2$Te$_3$(111) interaction potential can be described by a corrugated Morse potential with a well depth $D=(6.22\pm0.05)~\mathrm{meV}$, a stiffness $\kappa =(0.92\pm0.01)~\mathrm{\AA}^{-1}$ and a surface electronic corrugation of $(9.6\pm0.2)$\% of the lattice constant. 
The improved uncertainties of the atom-surface interaction potential should also enable the use in inelastic close-coupled calculations in order to eventually study the temperature dependence and the line width of selective adsorption resonances.
\end{abstract}

\end{frontmatter}

\section{Introduction}
\label{sec:intro}
Bi$_2$Te$_3$ is classified as a topological insulator (TI)\cite{Chen2009}, a class of materials which exhibit protected metallic surface states and an insulating bulk electronic structure\cite{Moore2010,Hasan2010,Zhang2011}. The modification of the electronic structure of topological surfaces upon adsorption of atoms and molecules has been subject to several studies\cite{Hsieh2009,Wray2011,Wang2015,Caputo2016}. However, the interaction of topological insulator surfaces with its environment, including the atom-surface interaction potential is largely unexplored by experiment.\\
Information about the detailed shape of the interaction potential can be gained from atom scattering experiments. In favourable cases the hard wall of the potential can be studied through the profile of the specular lattice rod, a form of interference in $\Delta k_{z}$\cite{Ellis2006,Ellis1995}. However, such methods are relatively insensitive to the form of the attractive interaction, which is the main concern in the present work. Here, we analyse observations of resonant scattering using $^3$He spin-echo spectroscopy measurements in combination with elastic close-coupling scattering calculations to determine the He--Bi$_2$Te$_3$(111) interaction potential. Atom-surface potentials can be measured to an extremely high accuracy by using selective adsorption resonances (SAR) in atom-surface scattering via the technique of $^3$He spin-echo spectrometry\cite{Jardine2004,Riley2007}.\\           
A detailed study of the atom-surface interaction on topological insulators is particularly interesting from a fundamental point of view. Precise measurements of atom-surface potentials offer a high-resolution window into the atom-surface interaction dynamics within the van der Waals regime, a field of intense theoretical interest in testing the ability of density functional theory calculations to simulate nonlocal interactions\cite{Brivio1999,Wu2001,Jean2004}. Hence the experimental data may assist in bench-marking of current theoretical approaches for the description of van der Waals forces. Such approaches become even more complicated for nanostructured surfaces\cite{Ambrosetti2016}. On topological insulator surfaces with their peculiar electronic surface effects our data and experimental approach may also help to understand the above mentioned influence of adsorption upon the electronic structure, where in particular the long-range part of the potential is responsible for band bending effects\cite{Forster2015}.\\
Selective adsorption phenomena appear in atom-molecule scattering off periodic surfaces due to the attractive part of the atom-surface interaction potential. According to Bragg's law, when an atom is scattered by a periodic surface, the change in the wavevector component parallel to the surface, $\mathbf{K}$, must be equal to a surface reciprocal lattice vector, $\mathbf{G}$. In the case of elastic scattering, the wavevector component perpendicular to the surface, $k_{f,z}$, is given via the conservation of energy and the kinematically-allowed $\mathbf{G}$-vectors for scattering are those for which $k_{f,z}^2$ is positive.\\
SARs occur when a He atom is diffracted into a channel which is kinematically disallowed ($k_{f,z}^2<0$) whilst simultaneously dropping into a bound state of the atom-surface potential. The kinematics of a SAR, involving a bound state of energy $- \left| \epsilon_n \right|$, is defined by the simultaneous conservation of energy and parallel momentum. The corresponding process can only take place if the difference between the energy of the incident atom and the kinetic energy of the atom moving parallel to the surface matches the binding energy $\epsilon_n$ of the adsorbed atom\cite{Hoinkes1992}:
\begin{equation}
E = \frac{\hbar^2 \mathbf{k}_i^2}{2m} = \frac{\hbar^2 (\mathbf{K}_i + \mathbf{G})^2}{2m} + \epsilon_n(\mathbf{K}_i,\mathbf{G}) .
\label{eq:kinematic}
\end{equation}
Since SARs correspond to the specific bound state energies $\epsilon_n$, of the He-surface interaction potential, the phenomenon provides a natural approach for studies of the atom-surface potential. It has only recently been shown that SARs in He scattering can even be used to reveal the degree of proton order in an ice surface\cite{Avidor2016}.\\
However, the majority of experimentally measured SARs is based on salts with the NaCl structure\cite{Hoinkes1992,Farias1998,Eichenauer1988,Benedek2001,Jardine2004,Riley2007,Debiossac2014} with some exceptions such as, adsorbate systems\cite{Kirsten1991,Riley2008}, stepped metal surfaces\cite{Apel1996,Farias1998} or the semimetal surfaces of Bi(111) and Sb(111)\cite{Kraus2013,Mayrhofer2013} and the semiconductor Si(111)-H\cite{Tuddenham2009}.

\section{Experimental Details}
\label{sec:experimental}
\begin{figure}[htb]
\centering
\includegraphics[width=0.48\textwidth]{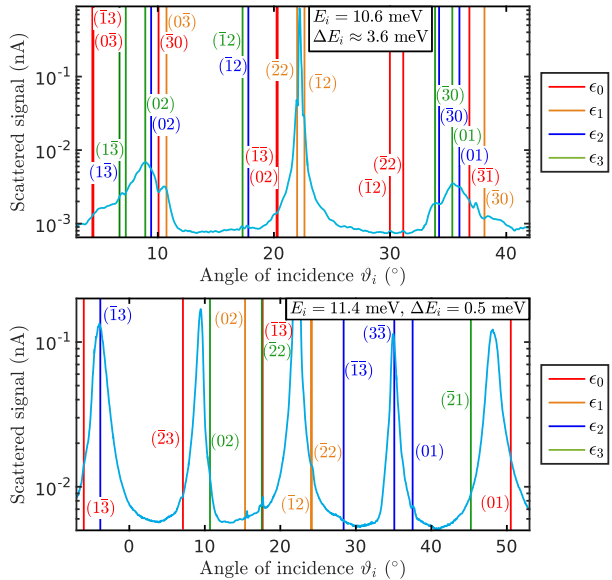}
\caption{Scattered He intensities (logarithmic scale) for Bi$_2$Te$_3$(111) versus incident angle $\vartheta_{i}$ along the {\GM} azimuth and at a sample temperature of 107 K. The central beam energy $E_i$ is 10.6 and 11.4 meV while the energy spread $\Delta E_i$ (full width at half maximum) is 3.6 and 0.5 meV in the top and bottom panels, respectively. In addition to the diffraction peaks, further peaks and dips corresponding to selective adsorption processes are identifiable. The vertical lines illustrate the kinematic conditions for four bound-state energies $\epsilon_0$-$\epsilon_3$. Each line is labelled with the Miller indices of the associated $\mathbf{G}$ vector for the resonance condition.}
\label{fig:DiffRes}
\end{figure}

\begin{figure}[htb]
\centering
\includegraphics[width=0.48\textwidth]{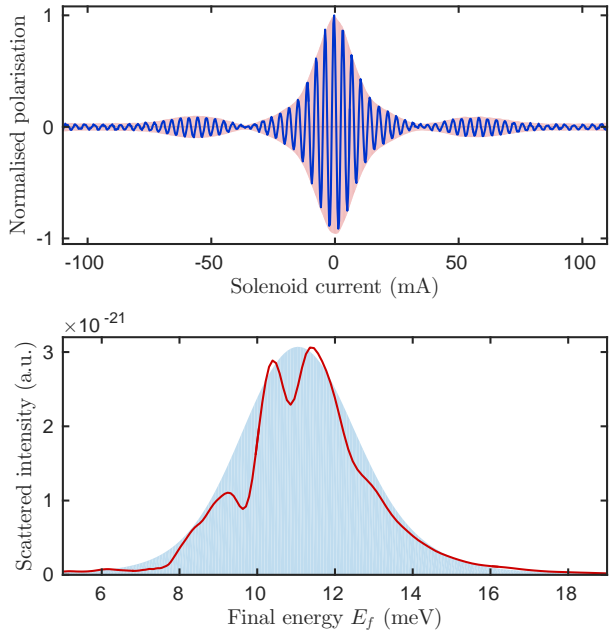}
\caption{Example of a typical measurement that includes selective adsorption resonances. The upper panel shows the detector signal for different currents in the incoming solenoid, which is proportional to the magnetic field. The overall decay envelope is due to the energy spread of the beam. The beating effect is an interference pattern due to the existence of multiple energy peaks within the overall beam energy distribution which are a consequence of SARs. The lower panel shows the signal after transformation to the energy scale (red solid line). The distribution of energy in the incident beam is approximately Gaussian and is illustrated by the shaded background. Resonances are evident in the difference between the line and the smooth background. Note that, in contrast to conventional experiments, the energy width of the resonances is given by their natural width and is much less than the energy spread in the beam.}
\label{fig:FTScan}
\end{figure}

In the present work we use $^3$He--Bi$_2$Te$_3$(111) selective adsorption data obtained with the Cambridge helium-3 spin-echo spectrometer\cite{Jardine2009}. A nearly monochromatic beam of $^3$He is scattered off the sample surface in a fixed 44.4$^{\circ}$ source-target-detector geometry. A detailed setup of the apparatus has been described in greater detail elsewhere\cite{Alexandrowicz2007,Jardine2009}.\\
Bi$_2$Te$_3$ exhibits a rhombohedral crystal structure which consists of quintuple layers bound to each other through weak van der Waals forces giving easy access to the (111) surface by cleavage\cite{Howard2013,Michiardi2014}(see Michiardi \emph{et al.}\cite{Michiardi2014} for details on the crystal growth procedure). The (111) cleavage plane is terminated by Te atoms with a hexagonal structure ($a=4.386~\mathrm{\AA}$)\cite{Tamtogl2017}. The Bi$_2$Te$_3$ single crystal used in the study was attached onto a sample holder using electrically and thermally conductive epoxy. The sample holder was then inserted into the chamber using a load-lock system\cite{Tamtogl2016a} and cleaved \textit{in-situ}. The sample holder can be heated using a radiative heating filament on the backside of the crystal or cooled down to $\approx 100~\mbox{K}$ using liquid nitrogen. The sample temperature was measured using a chromel-alumel thermocouple.\\
A measurement which can be used to identify SARs, is the so called $\vartheta$-scan, where the scattered beam intensity is measured as a function of the incident angle $\vartheta_i$, while the total scattering angle is fixed. In doing so, the momentum transfer parallel to the surface, given by $\lvert \Delta \mathbf{K} \rvert = \lvert \mathbf{k_i} \rvert ( \sin (\vartheta_f ) - \sin (\vartheta_i ) )$, is varied by changing the incident angle $\vartheta_i$. A typical diffraction scan for the {\GM} azimuth is shown in the lower panel of \autoref{fig:DiffRes}. In between the diffraction peaks, there may appear small peaks or dips in the scattered intensity which can be assigned to SARs, with the position of the peaks given by equation \eqref{eq:kinematic}. By changing the beam energy and the azimuthal angle, different SAR conditions can be met.\\
%############################################################################################################################################################
%############################################################################################################################################################
In \autoref{fig:DiffRes} rapid variations in scattered intensity have been identified with particular resonances. The resonance positions are indicated by vertical lines in \autoref{fig:DiffRes}, with annotations indicating the diffraction channel and bound-state index. In identifying particular resonances, we have assumed the free atom approximation, where the binding energy $\epsilon_{n}$, in \autoref{eq:kinematic}, is taken as a constant and is independent of $\mathbf{K}_{i}$ and $\mathbf{G}$. The approximation is valid in the limit of zero corrugation and is a useful starting point for the more detailed analysis performed below. A corrugated potential creates a more complex band-structure for the resonances where the dispersion is no longer that of a free-atom and nearby resonances may interact. Such effects are visible in a 3-D plot of scattered intensity against energy and azimuthal angle. The spin-echo spectrometer enables such plots from measurements of the specular intensity against azimuthal angle, as we describe below.
We will only briefly summarise the measurement of such a two-dimensional scan here, since a more detailed description can be found elsewhere\cite{Jardine2004}. We use the Fourier transform nature of the spectrometer to analyse the intensity distribution as a function of energy within the specularly scattered helium beam, to produce a complete data set of SARs for the particular scattering geometry used. Selective adsorption features were measured by performing a series of spin precession scans on the specularly scattered helium beam using one of the instruments spin-precession coils. The results were Fourier transformed onto an energy scale\cite{Jardine2009}, where selective adsorption processes appear as dips or peaks in the scattered intensity at specific characteristic energies.\\
To probe as many selective adsorption processes as possible, we used $^3$He nozzle conditions which gave a wide energy spread in the incident beam. Two sets of measurements were performed, one with the central beam energy at $8$ meV and a full width at half maximum of $1.66$ meV and one with the central beam energy around $11$ meV and a width of $3.80$ meV. \autoref{fig:FTScan} shows an example of such a scan with the raw and Fourier transformed data.\\
The spin precession measurements used precession solenoid currents between $-400$ and $+400$ mA in $0.33$ mA steps. The complete data set was built up from a series of 130 scans (65 at each nozzle temperature), taken at an azimuthal angle spacing of 0.5$^\circ$, to encompass the entire region within the \GM\ and \GK\ azimuthal directions. The scans at both beam energies were then combined into one plot. To further increase the visibility of bound state resonance features the scattered intensity was subtracted from the beam profile as given without the effect of bound state resonances. We use the average of the whole data set along the azimuthal direction as a measure of the beam profile\cite{Riley2008,Tuddenham2009} and the whole data set was then normalised after subtraction from the beam profile.\\
%############################################################################################################################################################
%############################################################################################################################################################
In Fig.\autoref{fig:AzimuthScan} the final data set is shown as a function of $\varphi$ (the azimuthal angle relative to the \GK\ azimuth) and $E_f$. A number of lines of high and low intensities, which we identify as SAR features, can be seen to run across the data set.\\

\section{Analysis}
For the atom-surface interaction we assume a corrugated Morse potential (CMP). Strictly speaking, the Morse potential does not exhibit the right $z^{-3}$ asymptotic behaviour and the long-range interaction may be described more accurately by modified versions of the interaction potential. However, as shown recently, Morse- or Morse-like potential functions are perfectly suitable for representing the bound state energies of semimetal surfaces\cite{Kraus2015,Kraus2014} and the use of the Morse potential allows to solve several steps within the close-coupling (CC) algorithm analytically, which greatly simplifies the computational cost.\\
For a three-dimensional atom-surface interaction potential the potential is written in terms of the lateral position $\mathbf{R}$ on the surface and the distance $z$ with respect to the surface:\cite{Armand1983}
\begin{equation}
    V(\mathbf{R},z) = D\left[\frac{1}{\nu_{0,0}} e^{-2\kappa\left[z-\xi(\mathbf{R})\right]}
    -2e^{-\kappa z}\right] 
    \label{eq:CMPpot}
\end{equation}
where $\kappa$ is the stiffness parameter, $D$ is the depth of the potential well and $\nu_{0,0}$ is the surface average of the exponent of the corrugation function. The electronic surface corrugation is given by $\xi(\mathbf{R})$ where $\mathbf{R}$ is the lateral position in the surface plane describing a periodically modulated surface with constant total electron density. $\xi(\mathbf{R})$ is described by the summation of cosine terms obtained from a Fourier series expansion based on a hexagonal unit cell\cite{Kraus2015,Mayrhofer2013}:
\footnotesize 
\begin{equation}\label{eq:corrugation}
\begin{split}
\xi(x,y) = \xi_0 & \left\{ \cos \left[ \frac{2\pi}{a} \left( x - \frac{y}{\sqrt{3}} \right) \right] + \cos \left[ \frac{2\pi}{a} \left( x + \frac{y}{\sqrt{3}} \right) \right] \right. \\
 & \quad \left. + \cos \left[ \frac{2\pi}{a} \frac{2y}{\sqrt{3}} \right] \right\} + h.o.
 \end{split}
\end{equation}
\normalsize
with $\xi_0$ the corrugation amplitude. The magnitude of the corrugation is typically given in terms of the peak-to-peak corrugation $\xi_{pp}$ of \eqref{eq:corrugation}.\\
The laterally averaged surface potential $V_0$ of \eqref{eq:CMPpot} is then given via:
\begin{equation}
    V_{0} (z) = D \left[ e^{-2\kappa z}-2e^{-\kappa z} \right]
\end{equation}
and the corresponding couplings can be found in\cite{Mayrhofer2013,Kraus2014,Kraus2015}. The bound states of the averaged potential are described by an analytical expression:
\begin{equation}
\epsilon_{n} = -D + \hbar \omega (n + 0.5)  \left(1 - \frac{n + 0.5}{2 \gamma}\right)
\label{eq:MP}
\end{equation}
with a positive integer $n$, $\omega = \kappa \sqrt\frac{2 D}{m}$ and $\gamma = \frac{2 D}{\hbar \omega}$, where $m$ is the mass of the impinging $^3$He atom.\\
In order to determine the best three-dimensional potential, we go through the following three-step process, starting with a laterally averaged atom-surface interaction potential followed by refining the three-dimensional potential via comparison of close-coupled calculations with the experimental data: 
\begin{enumerate}
    %\item ``Guess'' a laterally averaged atom-surface interaction potential based on the SAR positions and the free atom model.
    \item Obtain an approximate, surface-averaged potential using the free-atom model (\autoref{fig:DiffRes}).
    \item Determine the corrugation amplitude $\xi_0$ of the potential from diffraction measurements to acquire a three-dimensional potential.
    \item Simulate the experimental measurements using the corrugated potential and further improve the potential by comparison of the simulation with the experimental data.
\end{enumerate}
The process is described in detail below.

\subsection{Comparison with the free atom model}
\label{sec:comparefreeatom}
The free atom approximation (for parallel motion) assumes that the surface potential is adequately described simply by the laterally averaged component of the interaction potential. It corresponds to the case where the surface corrugation approaches zero, which is not possible in reality, since corrugation is necessary to provide the corresponding $\mathbf{G}$-vector for the resonance processes.\\
In particular for strongly corrugated systems the free atom approximation is no longer valid. Based on the diffraction peak intensities a peak-to-peak corrugation of 9\% of the surface lattice constant was found for Bi$_2$Te$_3$\cite{Tamtogl2017} and about 5\% in the case of Bi(111)\cite{Kraus2015}. Though still smaller than the corrugation of several semiconductor and insulating surfaces\cite{Farias1998,Tamtogl2017} these are rather large values. Hence one would expect that band structure effects play a significant role.\\
Nevertheless, despite its limitations and simplifying nature, the free atom approximation provides a good starting point to understand selective adsorption phenomena. To calculate the positions, the kinematic condition from \eqref{eq:kinematic} is written in terms of the incident angle $\vartheta_i$ and the incident wave vector $k_i$, which corresponds to the beam energy as well as the components of the scattering vector $\mathbf{G}~=~(G_\parallel, G_\perp)$:
\begin{equation}
-\cos^2(\vartheta_i) k_i^2 + 2 \sin(\vartheta_i)G_\parallel k_i +G_\parallel^2 +G_\perp^2 -\frac{2m}{\hbar^2}|\epsilon_n| = 0 
\label{eq:kinematic2}
\end{equation}
Here $\mathbf{G}$ is split into the components $G_\parallel$ and $G_\perp$  parallel and normal to the incidence plane, respectively. Diffraction scans with some SARs are useful in order to obtain a first idea about the bound state energies: Solving \eqref{eq:kinematic2} provides an estimate of the bound state energy $\epsilon_n$ associated with a peak or dip at a certain $\vartheta_i$ in the diffraction scan.\\
%############################################################################################################################################################
\begin{figure}[!ht]
\centering
\subfloat[Measured specular scattering intensity as a function of $\varphi$ and $E_f$, for $^3$He scattering from Bi$_2$Te$_3$(111) surface, where the colour scale corresponds to the intensity. The crystal was held at 110 K during the measurement and the underlying beam profile of the incoming He beam has been removed to obtain a spectrum of the bound state resonances. A number of lines of high and low intensities, which we identify as selective adsorption resonances features, can be seen to run across the data set.\label{fig:AzimuthScan}]{\includegraphics[width=0.45\textwidth]{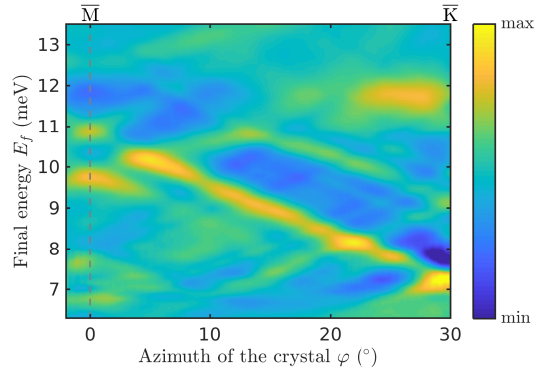}}\\
\subfloat[Contour plot of the simulated data set based on the optimised three-dimensional parameters. To obtain the same contour plot as in the experiment (Fig.\autoref{fig:AzimuthScan}) the  beam profile is included and a small Gaussian blur is introduced which accounts for inelastic effects that are present in the experimental data (measurement at finite temperature). \label{fig:AzimuthSimFinal}]{\includegraphics[width=0.45\textwidth]{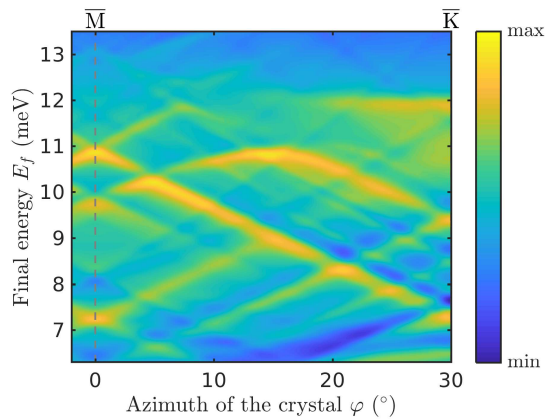}}
\caption{Experimental data set as a function of $(\varphi , E_f )$ together with the simulated data set based on the optimised atom-surface interaction potential.}
\label{fig:Azimuth}
\end{figure}
%############################################################################################################################################################
The lower panel of \autoref{fig:DiffRes} shows a diffraction scan for the {\GM} azimuth with an incident beam energy of $11.4~\mbox{meV}$. A couple of small peaks and some shoulders at the diffraction peak positions, which may be caused by SARs, are visible in the scan. The upper panel of \autoref{fig:DiffRes} shows a diffraction scan along {\GM} with a wide energy spread giving rise to the much broader diffraction peaks. SARs sitting on the diffraction peak positions are now much more evident, however, it complicates the analysis since the incoming beam energy $E_i$ and consequently $k_i$ in \autoref{eq:kinematic2} is no longer clearly defined.\\
Nevertheless, we can use the positions of the peaks and dips in \autoref{fig:DiffRes} to get a first idea about the bound state energies $\epsilon_n$ and the associated laterally averaged potential. The vertical lines in \autoref{fig:DiffRes} display the SAR conditions based on \autoref{eq:kinematic2} for four bound state energies $\epsilon_0$-$\epsilon_3$ and the reciprocal lattice vectors $\mathbf{G}$ as labelled in the graph. Based on these SAR features, there appear to be four bound state energies with $\epsilon_0 \approx 4.5~\mbox{meV}$, $\epsilon_1 \approx 1.9~\mbox{meV}$, $\epsilon_2 \approx 0.3~\mbox{meV}$ and $\epsilon_3 \approx 0.05~\mbox{meV}$.\\
The bound state energies $\epsilon_n$ of the laterally averaged potential can be calculated analytically using \eqref{eq:MP}. Using an optimisation routine based on the four bound state energies, we obtain a potential with the parameters $D=(6.4\pm0.3)~\mathrm{meV}$ and $\kappa =(0.94\pm0.06)~\mathrm{\AA}^{-1}$

\subsection{Comparison with elastic close-coupled calculations}
\label{sec:comparecalcs}
\begin{figure}[!ht]
\centering
\subfloat[\label{fig:AzimuthShiftSim}Close coupled calculations of the $^3$He--Bi$_2$Te$_3$(111) specular scattering intensity for the same conditions as the experimental data, using the potential parameters obtained in \ref{sec:comparefreeatom}. The dotted and dash-dotted lines show a number of kinematic conditions (\autoref{eq:kinematic2}) based on the laterally averaged potential.]{\includegraphics[width=0.45\textwidth]{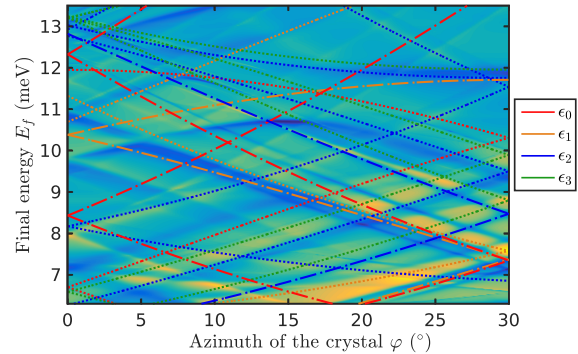}}\\
\subfloat[\label{fig:AzimuthShiftExp}Experimental data from Fig.\autoref{fig:AzimuthScan} with several kinematic conditions (\autoref{eq:kinematic2}) of the laterally averaged potential plotted as dash-dotted lines. ]{\includegraphics[width=0.45\textwidth]{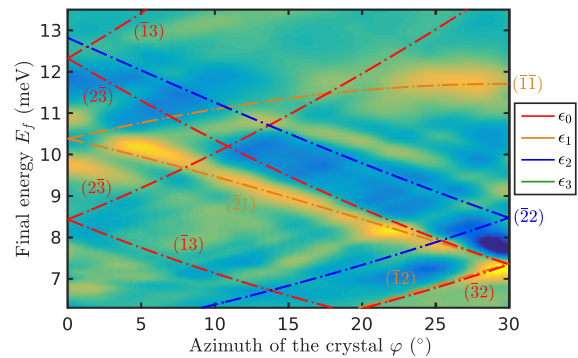}}
\caption{(Same colormap as \autoref{fig:Azimuth}) A comparison of the resonance positions visible in the simulated data with the kinematic conditions of the laterally averaged in (a) illustrates that band structure effects play a significant role, giving rise to strong deviations. In order to further refine the three-dimensional potential we have picked a number of specific resonances of which the kinematic conditions are illustrated as dash-dotted lines. We then minimise the deviation in terms of the $(\varphi,E_f)$ position of these specific resonances between the experimental data in (b) and the simulation in (a) to obtain a best-fit three-dimensional potential.}
\label{fig:AzimuthOffset}
\end{figure}

While in the free atom approximation the coupling term vanishes, the band structure diagram in such a situation would consist entirely of parabolic bands. For strongly corrugated systems, contributions of the higher-order Fourier components in the surface potential become significant and can no longer be neglected. Hence resonance positions calculated using a corrugated surface potential give rise to a substantial deviation from the free atom parabolic bands in analogy to the occurrence of energy gaps in the electronic band structure at the Brillouin zone boundary. Similarly a splitting of parabolic bands and the development of energy gaps at zones of degeneracy may occur due to the spatial periodicity of the atom-surface interaction potential. These effects have been highlighted in the past\cite{Chow1976,Hoinkes1980,Manson1983,Vargas1996,Subramanian1999} and an exact description of the measured data is only possible by exact quantum mechanical calculations based on the three-dimensional potential.\\
In a purely elastic scattering scheme, scattering of a He atom with incident wavevector $\mathbf{k}_i$, is described by the time-independent Schr\"{o}dinger equation with the potential as given by \autoref{eq:CMPpot}. Together with a Fourier expansion of the wave function it gives rise to a set of coupled equations for the diffracted waves which are solved for in the close-coupling algorithm using finite set of closed channels\cite{Sanz2007,Kraus2013}.\\
In a first step the corrugation amplitude of the three-dimensional potential needs to be determined. Therefore the elastic peak intensities are simulated using the close-coupling algorithm, starting with the parameters of the laterally averaged potential obtained in the previous section. The calculated purely elastic intensities are corrected with the Debye-Waller attenuation and compared with the experimentally determined peak areas\cite{Tamtogl2017}. The peak-to-peak corrugation $\xi_{pp}$ was varied, over a range of $0.005 - 0.7~\mbox{\AA}$ with a step width of $0.005~\mbox{\AA}$ giving rise to a best fit with the experimentally determined peak areas at $\xi_{pp} = 0.45~\mbox{\AA}$.\\
Once a starting point for the parameters of the three-dimensional potential \eqref{eq:CMPpot} is known, the close-coupling algorithm is used to calculate a data-set similar to the measurement in Fig.\autoref{fig:AzimuthScan}. Therefore the elastically scattered intensity for the secular scattering condition is calculated for a set of different beam energies and for different azimuthal orientations of the crystal. The scattered intensity for each set of $(\varphi,E_f )$ is shown in the contour plot of Fig.\autoref{fig:AzimuthShiftSim}.\\
%############################################################################################################################################################
%############################################################################################################################################################
Several SARs features appear either as local maxima or minima in the contour plot. The dotted and dash-dotted lines which are superimposed onto the contour plot illustrate a number of kinematic conditions, \autoref{eq:kinematic2}, based on the laterally averaged potential. Note that in regions where there are several resonances based on the free atom model, it is not always possible to identify a clear line shape in the simulated data. In particular, several of the kinematic conditions show a strong deviation with respect to the lines running through the simulated data. There are also several lines in the simulated data which are not matched by any of the kinematic conditions. It illustrates that the above mentioned band structure effects play a significant role and hence the coupling between the scattering channels as present in the close-coupling simulation is essential\cite{Sanz2007}.\\
Since the deviations of the kinematic conditions from the full quantum-mechanical calculation depend strongly on the potential parameters, the best three-dimensional potential is hard to find. Furthermore, the kinematic conditions do not predict whether a resonance condition gives rise to a maximum or minimum. Therefore, a self-consistency cycle or a method which does not use the kinematic condition has to be evaluated to avoid this problem. The latter can be achieved by comparing the SAR positions obtained from the close-coupling calculation directly to the experimental data, which requires the simulation of data sets for a high number of parameter sets. To make this option viable, the search space has to be reduced, so that it can be scanned in a reasonable amount of time. Therefore, we start with the potential found in \ref{sec:comparefreeatom} as the centre of the parameter space and create a parameter grid around it, which reduces the number of simulations and enables the application of parallel computation. In the case of the corrugated Morse potential, the parameter space can be reduced to two dimensions since the corrugation $\xi_{pp}$ can be determined beforehand via comparison with the experimentally determined peak areas.\\
To quantify the quality of the resulting simulation, a $\chi^2$-test is used, testing that the positions $(\varphi,E_f)$ of the SARs in the simulated data set coincide with the positions in the experimental data set. It leads to an equation for the $\chi^2$ sum, which adds the squares of the difference between the position of the resonances from the simulation $\epsilon_{i,s}$ and the position seen in the experimental data $\epsilon_{i,m}$ divided by the sum of the standard deviations of the experiment $\sigma_{i,m}$ and the simulated data $\sigma_{i,s}$:
\begin{equation}
   \chi^2 = \sum_i \left(
   \frac{\epsilon_{i,s}-\epsilon_{i,m}}{\sigma_{i,m}+\sigma_{i,s}}\right)^2
\label{eq:chi2}
\end{equation}
Here we assume that the resonance positions follow a normal distribution, since they were measured by hand using the image analysis tool called Fiji\cite{FijiPaper} on a graphical representation of the simulation. To obtain the standard deviation, a cut at a fixed azimuthal angle $\varphi$ was taken and the half width at half maximum of the resonance signal was used, after subtracting the background.\\
The calculation of the $\chi^2$ value is done on a grid with the potential depth $D$ spanning from $5.9$ to $6.4$ meV with a step width of $0.05$ meV and the potential stiffness $\kappa$  spanning from $0.88$ to $1.00$ \AA$^{-1}$ with a step width of $0.005$ \AA$^{-1}$ resulting in 143 potentials. For the calculation of the cost function we have used three resonances, associated with $\epsilon_0$, $\epsilon_1$ and $\epsilon_2$ which is illustrated in \autoref{fig:AzimuthOffset}: The kinematic conditions for these resonances are illustrated as dash-dotted lines on top of the experimental as well as the simulated data.\\
%############################################################################################################################################################
\begin{figure}[htb]
\centering
\includegraphics[width=0.4\textwidth]{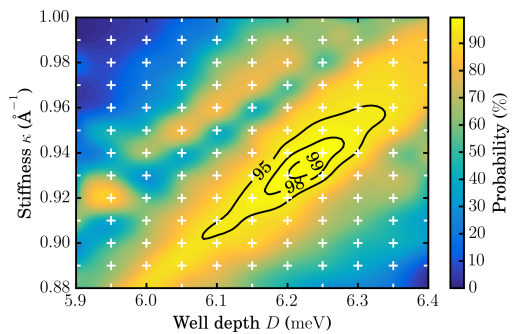}
\caption{The probability that the three resonances $\epsilon_0$ - $\epsilon_2$, associated with specific reciprocal lattice vectors (see \autoref{fig:AzimuthOffset}) appear at the same position in the experimental and the simulated data. The white crosses show the evaluated points on the parameter grid and the contour lines show regions for $\alpha$-values of 1\%, 2\% and 5\% (corresponding to a confidence interval of 99\%, 98\% and 95\%, respectively).}
\label{fig:chigrid}
\end{figure}
%############################################################################################################################################################
The result of the optimisation is plotted in \autoref{fig:chigrid} as a colour-map plot, showing the regions for three different significance levels $\alpha$ of 1\%, 2\% and 5\% (corresponding to a confidence interval of 99\%, 98\% and 95\%), respectively.\\
Once the best-fit potential parameters $D$ and $\kappa$ have been found, the corrugation $\xi_{pp}$ is further refined using again a comparison with the experimentally determined diffraction peak areas. Following this approach, the parameters of the best-fit three-dimensional potential based on a significance level of $\alpha = 2\%$ are:
\begin{gather*}
D=(6.22\pm0.05)~\mathrm{meV}\\
\kappa =(0.92\pm0.01)~\mathrm{\AA}^{-1}\\
\xi_{pp} = (0.42\pm0.01)~\mathrm{\AA}
\end{gather*}
Compared to the results from \ref{sec:comparefreeatom}, the well depth $D$ and the corrugation are now somewhat smaller while the stiffness $\kappa$ increased. While the well depth and stiffness obtained from the free particle model may be used as a reasonable estimate, the uncertainties of all three potential parameters are significantly reduced by comparison with the close-coupling calculations. More importantly, the free particle model can only provide an estimate for the position of the resonances but cannot reproduce the shape of the resonances, in particular whether there appear maxima or minima, which is inherently obtained from the close-coupling calculations.\\
A simulated data set based on the same conditions as the experimental data set with the optimised three-dimensional parameters is shown in Fig.\autoref{fig:AzimuthSimFinal}. 
To obtain the same contour plot as in the experiment (Fig.\autoref{fig:AzimuthScan}) the beam profile is included: The simulated data is first multiplied with the beam profile to account for the energy distribution of the incoming He beam after which the data is again subtracted from the beam profile to follow the same procedure as for the treatment of the experimental data. Finally a Gaussian blur with a standard deviation of 80 $\upmu$eV in energy is introduced. The blur is a measure of the average linewidth of the resonances and accounts for the fact that our purely elastic analysis with the corrugated Morse potential fails to reproduce the linewidths of the resonances as measured in the experiment. Several factors may contribute to a resonance linewidth\cite{Tuddenham2009} including inelasticity, disorder and the distribution of the corrugation between the attractive and repulsive parts of the potential. A comparison of the simulated data including the Gaussian blur with the experimental data (\autoref{fig:Azimuth}) shows that all main features are very well reproduced and appear at the right position in terms of $\varphi$ and $E_f$.\\
For a rough estimate of the potential depth $D$, the ratio between the potential depth and the average atomic mass of the sample can be used. It gives rise to a value of $0.039~\mathrm{meV~  u}^{-1}$ for the Bi$_2$Te$_3$(111) surface, which is in good agreement with similar material surfaces such as Sb(111) ($0.035~\mathrm{meV~  u}^{-1}$)\cite{Kraus2014} and Bi(111) ($0.038~\mathrm{meV~  u}^{-1}$)\cite{Kraus2015}. The value of the well depth $D$ itself, is between those found for Sb(111)($4.3~\mbox{meV}$)\cite{Kraus2014} and Bi(111) ($7.9~\mbox{meV}$)\cite{Kraus2015} while being considerably lower than the one found for graphite(0001)($\approx 16$ meV)\cite{Tamtogl2015,Hoinkes1980}.\\
The stiffness $\kappa$ of the He--Bi$_2$Te$_3$(111) potential is much larger compared to the He--Sb(111) potential ($0.39~\mbox{\AA}^{-1}$)\cite{Kraus2013} and indeed rather comparable to the He--LiF(001) potential\cite{Hoinkes1992}. On the one hand this could be connected with the insulating interior and polarisability of the topological material. On the other hand the He--Bi(111) potential has a similar stiffness ($0.88~\mbox{\AA}^{-1}$)\cite{Kraus2015} and it seems to be difficult to identify a general trend based on the stiffness $\kappa$.\\
The peak-to-peak corrugation of the final optimised potential is  $(9.6\pm0.2)\%$ relative to the lattice constant and hence only slightly larger compared to a first analysis based on a rough estimate of the potential\cite{Tamtogl2017}. This surface electronic corrugation is larger than the ones found for low-index metal surfaces\cite{Farias1998,Tamtogl2015} while being similar to the corrugation of semimetals such as Bi(111)(5\%)\cite{Kraus2015}, graphite(0001)(8.6\%)\cite{Boato1978,Boato1979} and Sb(111)($\approx15\%$)\cite{Kraus2014}.\\
Finally, inelastic processes and phonon mediated SARs have been identified in experiments and proven to play important roles\cite{Farias1998,Hoinkes1992}, also for similar systems as in our study, e.g. for helium scattering of the Bi(111) surface\cite{Kraus2013}. However, from a theoretical point of view, these effects have been mainly considered in the limit of low corrugated surfaces\cite{MiretArtes1995,Brenig2004,Sibner2008,MartinezCasado2014}. Since the inelastic scattering amplitudes involving bound states depend sensitively on both the repulsive and attractive parts of the potential they provide a discriminating test of the atom-surface interaction potential and we hope that our work will initiate further theoretical investigations in this direction.\\

\section*{Summary and Conclusion}
In summary, we have determined an atom-surface interaction potential for the He--Bi$_2$Te$_3$(111) system by analysing selective adsorption resonances. Following an initial free-particle model analysis, we use elastic close-coupling calculations to obtain an exact three-dimensional potential based on ultrahigh resolution $^3$He spin-echo spectroscopy measurements. Based on this analysis, the He--Bi$_2$Te$_3$(111) interaction potential is best described by a corrugated Morse potential with a well depth $D=(6.22\pm0.05)~\mathrm{meV}$, a stiffness $\kappa =(0.92\pm0.01)~\mathrm{\AA}^{-1}$ and a surface electronic corrugation of $(9.6\pm0.2)\%$ of the lattice constant.\\
To our knowledge, this work describes for the first time the determination of a high precision empirical atom-surface interaction potential of a topological insulator. The potential found in our study may assist in the development of first-principles theory where van der Waals dispersion forces play an important role and the improved uncertainties of the potential should also enable the use in inelastic close-coupled calculations. While the calculation of the scattered intensities including inelastic resonances requires the numerical solution of a large set of close-coupling equations which must be sufficiently large to assure convergence, with an exact potential at hand this should eventually allow to study the temperature dependence and the line width of selective adsorption resonances.

\section*{Acknowledgement}
Upon his retirement from Freie Universit\"at Berlin, Karl-Heinz Rieder enabled the transfer of his last He atom scattering machine to Graz. W. E. E. and the Graz group are grateful for his encouragement to start the investigation of semimetal surfaces which later broadened towards the class of topological materials.\\
We would like to thank P. Kraus for many helpful discussions. One of us (A.T.) acknowledges financial support provided by the FWF (Austrian Science Fund) within the project J3479-N20. The authors gratefully acknowledge support by the FWF within the project P29641-N36 and financial support by the Aarhus University Research Foundation, VILLUM FONDEN via the Centre of Excellence for Dirac Materials (Grant No. 11744) and the SPP1666 of the DFG (Grant No. HO 5150/1-2). E.M.J.H. and B.B.I. acknowledge financial support from the Center of Materials Crystallography (CMC) and the Danish National Research Foundation (DNRF93). S. M.-A. is grateful for financial support by a grant with Ref. FIS2014-52172-C2-1-P from the Ministerio de Econom\'ia y Competitividad (Spain).

%%%%%%%%%%%%%%%%%%%%%%%%%%%%%%%%%%%%%%%%%%%%%%%%%%%%%%%%%%%%%%%%%%%%%
\section*{References}
\footnotesize{
\bibliography{literature}
}

\clearpage
\begin{figure*}
\centering
\includegraphics[width=0.5\textwidth]{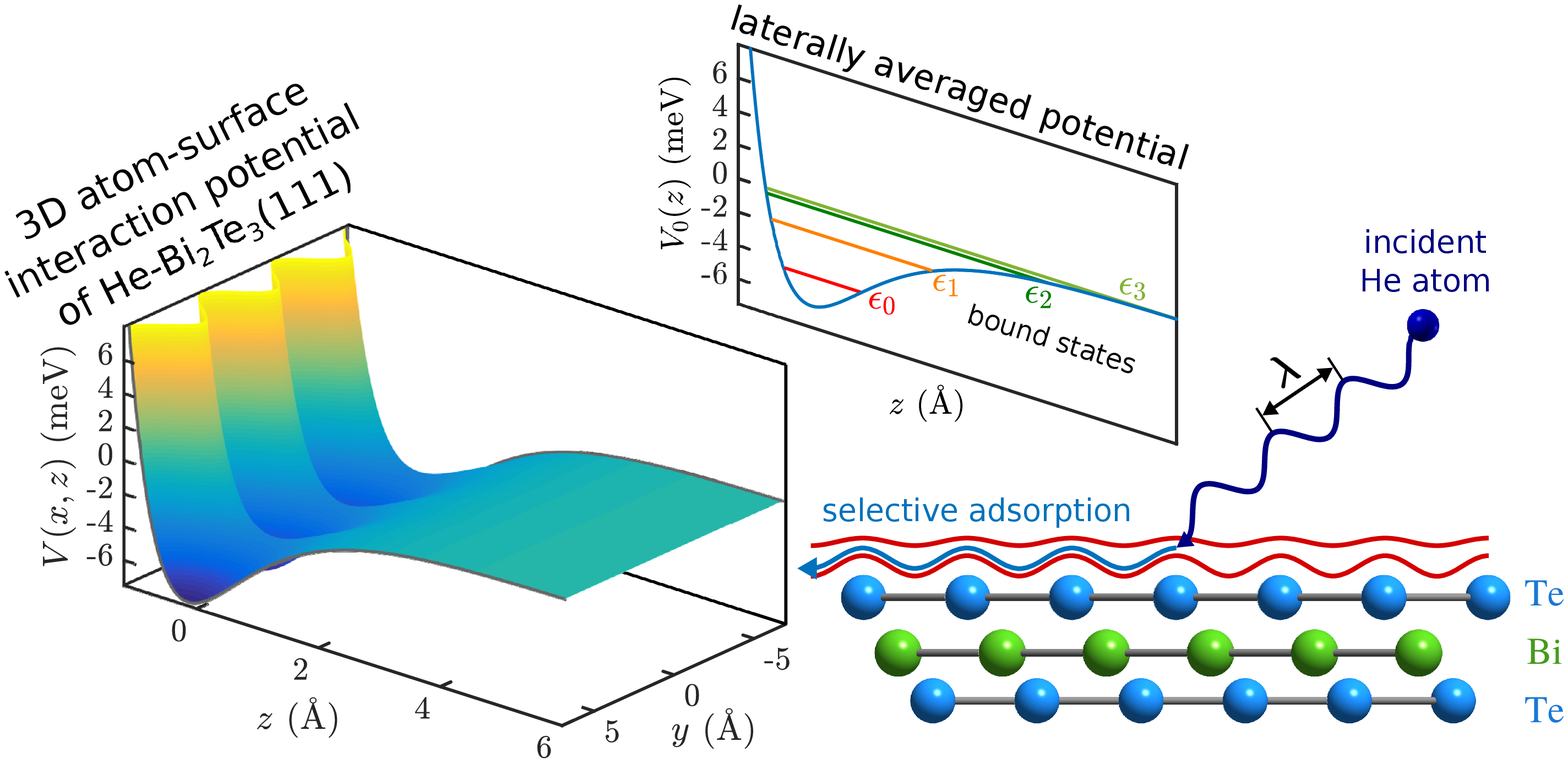}
\caption{Graphical Abstract}
\end{figure*}

\end{document}